\documentclass[fleqn,usenatbib]{mnras}

\usepackage{newtxtext,newtxmath}

\usepackage[T1]{fontenc}

\DeclareRobustCommand{\VAN}[3]{#2}
\let\VANthebibliography\thebibliography
\def\thebibliography{\DeclareRobustCommand{\VAN}[3]{##3}\VANthebibliography}

\usepackage{graphicx}	
\usepackage{adjustbox}  
\usepackage{amsmath}	
\usepackage{wasysym}    
\usepackage{bm}         
\usepackage{xspace}
\usepackage{soul}
\usepackage{booktabs} 
\usepackage{stfloats}
\makeatletter 
  \patchcmd{\NAT@citex}
    {\@citea\NAT@hyper@{%
      \NAT@nmfmt{\NAT@nm}%
      \hyper@natlinkbreak{\NAT@aysep\NAT@spacechar}{\@citeb\@extra@b@citeb}%
      \NAT@date}}
    {\@citea\NAT@nmfmt{\NAT@nm}%
    \NAT@aysep\NAT@spacechar\NAT@hyper@{\NAT@date}}{}{}

  \patchcmd{\NAT@citex}
    {\@citea\NAT@hyper@{%
      \NAT@nmfmt{\NAT@nm}%
      \hyper@natlinkbreak{\NAT@spacechar\NAT@@open\if*#1*\else#1\NAT@spacechar\fi}%
        {\@citeb\@extra@b@citeb}%
      \NAT@date}}
    {\@citea\NAT@nmfmt{\NAT@nm}%
    \NAT@spacechar\NAT@@open\if*#1*\else#1\NAT@spacechar\fi\NAT@hyper@{\NAT@date}}
    {}{}
\makeatother

\newcommand\Msun{\text{M}_{\astrosun}} 
\newcommand\orcid[1]{\href{http://orcid.org/#1}{\adjustbox{trim={-.15\width} {0\height} {-.15\width} {0\height},clip}{\includegraphics[height=10pt]{orcid.pdf}}}}
\newcommand\LCDM{\Lambda \text{CDM}}

\title[Constraining IDE models with halo $c$--$M$ relation]{Constraining interacting dark energy models with the halo concentration -- mass relation}

\author[Y. Zhao et al.]{
Yu Zhao,$^{1}$\thanks{E-mail: yuzhao0108@gmail.com}
Yun Liu,$^{2,3}$\thanks{E-mail: liuyun@nao.cas.cn}
Shihong Liao,$^{4}$
Jiajun Zhang,$^{5}$\thanks{E-mail: jjzhang@shao.ac.cn}
Xiangkun Liu$^{6}$
and Wei Du$^{7}$
\\
$^{1}$Department of Physics, Liaoning University, Shenyang 110036, China\\
$^{2}$National Astronomical Observatories, Chinese Academy of Sciences, Beijing 100101, China\\
$^{3}$University of Chinese Academy of Sciences, Beijing 100049, China\\
$^{4}$Department of Physics, University of Helsinki, Gustaf Hällströmin katu 2, FI-00014 Helsinki, Finland\\
$^{5}$Shanghai Astronomical Observatory, Chinese Academy of Sciences, Shanghai 200030, China\\
$^{6}$South-Western Institute for Astronomy Research, Yunnan University, Kunming 650500, China\\
$^{7}$Shanghai Key Lab for Astrophysics, Shanghai Normal University, Shanghai 200234, China
}

\date{Accepted XXX. Received YYY; in original form ZZZ}

\pubyear{2022}

\begin{document}
\label{firstpage}
\pagerange{\pageref{firstpage}--\pageref{lastpage}}
\maketitle

\begin{abstract}
The interacting dark energy (IDE) model is a promising alternative cosmological model which has the potential to solve the fine-tuning and coincidence problems by considering the interaction between dark matter and dark energy. Previous studies have shown that the energy exchange between the dark sectors in this model can significantly affect the dark matter halo properties. 
In this study, utilising a large set of cosmological $N$-body simulations, we
analyse the redshift evolution of the halo concentration--mass ($c$--$M$) relation in the IDE model, and show that the $c$--$M$ relation is a sensitive proxy of the interaction strength parameter $\xi_2$, especially at lower redshifts. Furthermore, we construct parametrized formulae to quantify the dependence of the $c$--$M$ relation on $\xi_2$ at redshifts ranging from $z=0$ to $0.6$. Our parametrized formulae provide a useful tool in constraining $\xi_2$ with the observational $c$--$M$ relation. As a first attempt, we use the data from X-ray, gravitational lensing, and galaxy rotational curve observations and obtain a tight constraint on $\xi_2$, i.e. $\xi_2 = 0.071 \pm 0.034$. 
Our work demonstrates that the halo $c$--$M$ relation, which reflects the halo assembly history, is a powerful probe to constrain the IDE model.
\end{abstract}



\section{Introduction}

The Lambda--cold dark matter ($\Lambda$CDM) model, which assumes two independently evolving dark sectors (i.e. the dark energy cosmological constant $\Lambda$ and the cold dark matter) as the major energy budgets in the Universe, has been spectacularly successful in explaining a large body of observations such as the cosmic microwave background \citep[CMB,][]{Hinshaw2013,Planck2020}, the large-scale structures \citep{Alam2021}, and the Type Ia supernovae observations \citep{Riess1998,Perlmutter1999}. It also provides a basic theoretical framework for modern galaxy formation models \citep{Benson2010,Somerville2015}. However, there are still critical challenges for this standard cosmological model in both theoretical and observational sides. The nature of both dark energy (DE) and dark matter (DM) is still an open question. For the cosmological constant, the long-standing fine-tuning and coincidence problems \citep{Weinberg1989,Zlatev1999} remain unsolved. For cold dark matter, there are discrepancies between the theoretical predictions and the observations at small-scales which have been summarised as the missing satellite problem \citep{Klypin1999,Moore1999}, the core-cusp problem \citep{Flores1994,Moore1994}, and the too-big-to-fail problem \citep{Boylan-Kolchin2011}; see \citet{Bullock2017} for a review. Last but not least, as the precision of cosmological observations improves, there are growing tensions among the measurements of cosmological parameters from different observational probes, such as the Hubble tension \citep{Riess2020,Wong2020,Riess2022} and the $\sigma_8$ tension \citep{Hikage2019,Hildebrandt2020,Asgari2021,Amon2022}.

Given these challenges and tensions in the $\Lambda$CDM model, there have been extensive attempts to introduce/study alternative models in the community. One of the alternative proposals is the so-called interacting dark energy (IDE) model \citep[see][for a review]{Wang2016}. Compared to the $\Lambda$CDM model, the IDE models assume that there are extra interactions between DM and DE and phenomenologically there is an energy exchange between these two dark sectors. Usually, the interaction kernels in phenomenological IDE models are assumed to be proportional to the linear combination of the DM and DE energy densities. It has been shown that IDE models can potentially explain the fine-tuning and coincidence problems \citep[e.g.][]{Amendola2000,Amendola2003,Pavon2005,Amendola2006,Olivares2006,Bohmer2008,Chen2008,delCampo2008} and alleviate the Hubble tension and the $\sigma_8$ tension \citep[e.g.][]{Costa2017,Ferreira2017,An2018,DiValentino2020}. This has made the IDE models a very promising beyond-$\Lambda$CDM candidate.

Until recently, most of the constraints on IDE models have come from large-scale and high-redshift observations, e.g. cosmic expansion history, CMB, and large-scale structures \citep[e.g.][]{Costa2017,Ferreira2017,An2018}. These observations can put tight constraints on the IDE models in which the interaction kernels depend on the DM density, e.g. the so-called IDE3 and IDE4 models in \citet{Wang2016}. For the IDE models in which the interaction kernels are proportional to the DE density, as the DE component only becomes dominant at the late epoch of the Universe, we expect that the DM-DE interactions in these models become more significant at low redshifts and have remarkable impacts on the non-linear structure formation (e.g. the DM halo properties). In this study, we aim to use small-scale observations as a complementary probe to help to constrain the IDE models in which the interaction kernels are proportional to the DE density, i.e. the so-called IDE1 (in which DM decays into DE) and IDE2 (in which DE transfers energy into DM) models in \citet{Wang2016}.

Cosmological $N$-body simulations are an essential tool in solving the non-linear structure formation process \citep{Frenk2012,Angulo2022}. Recently, there have been attempts to simulate the structure formation in IDE models \citep[e.g.][]{Baldi2010,Baldi2011,Baldi2012,Carlesi2014,Penzo2016,LHuillier2017,Zhang2018}. Among these works, \citet{Zhang2018} developed a fully self-consistent numerical simulation pipeline for IDE models, \textsc{me-gadget}, which is based on the widely used \textsc{gadget-2} code \citep{Springel2005}. This new simulation pipeline enabled \citet{Zhang2019} to simulate the non-linear structure formation to redshift $z=0$ and constrain the interaction between DM and DE by comparing the simulations with the SDSS galaxy-galaxy weak lensing measurements. Compared to previous studies, they found tighter constraints on the interaction strength parameter for the IDE1 and IDE2 models. With the same pipeline, \citet{Liu2022} performed a systematic study on the halo properties (e.g. halo growth histories, density profiles, spins, and shapes) in IDE models. They showed that compared to the $\Lambda$CDM haloes, haloes in the IDE models can have significantly different formation histories and properties. For example, in the IDE1 model, the structure formation is markedly slowed down and haloes have lower masses and looser internal density distributions. Most interestingly, they found that the halo concentration--mass relation ($c$--$M$ relation) is particularly sensitive to the interaction strength parameter, suggesting that the $c$--$M$ relation can be a powerful probe to constrain IDE models.

The halo concentration quantifies the matter distribution inside a virialized halo and depends on the halo accretion history \citep{Wechsler2002,Zhao2003,Zhao2009}. Thus, the relation between halo concentrations and masses (the so-called $c$--$M$ relation) reflects the formation history of DM haloes and carries important cosmological information. Given its significance in cosmology, the $c$--$M$ relation has been extensively studied in $\Lambda$CDM simulations \citep[e.g.][]{Bullock2001,Gao2008,Prada2012,Dutton2014,Ludlow2014,Child2018,Wang2020} and observations \citep[e.g.][]{Mandelbaum2006,Comerford2007,Gastaldello2007,Ettori2010,Sereno2010,Martinsson2013,Du2014,Meneghetti2014,Du2015,Merten2015,Amodeo2016,Groener2016}.

Following the study of the IDE $c$--$M$ relations in \citet{Liu2022}, in this work, we employ a set of high-resolution cosmological $N$-body simulations, which covers a larger IDE model parameter space, to further explore the dependence of IDE $c$--$M$ relations on the interaction strength parameter $\xi_2$, and attempt to use the observational $c$--$M$ relation to constrain $\xi_2$.

This paper is organized as follows. In Section~\ref{sec:2}, we describe the phenomenological IDE models, and the IDE and $\Lambda$CDM simulations. In Section~\ref{sec:3}, we present the computations of the $c$--$M$ relation from simulations, and analyse the parametrized $c$--$M$ relation as the function of redshift and $\xi_2$. We describe the observational data and use them to constrain the IDE interaction parameter in Section~\ref{sec:4}. Finally, we summarise and discuss our results in Section~\ref{sec:5}.

\section{Methods}
\label{sec:2}
\subsection{Phenomenological IDE models}
\label{sec:2.1}

\begin{table*}
	\centering
	\begin{tabular}{cccccccc}
		\toprule  
		Parameter&$\LCDM$&IDE1\_A&IDE1\_B&IDE1\_C&IDE1\_D&IDE2\_A&IDE2\_B\\ 
		\midrule  
		$\Omega_{\rm b}h^2$&0.02225&0.02223&0.022237&0.022237&0.022243&0.022325&0.02224\\
		$\Omega_{\rm c}h^2$&0.1198&0.0792&0.0927&0.0995&0.1063&0.12745&0.1351 \\
		$\ln{10^{10} A_{\rm s}}$&3.094&3.099&3.099&3.0965&3.097&3.0955&3.097\\
		$n_{\rm s}$&0.9645&0.9645&0.9645&0.9645&0.9643&0.9644&0.9643\\
		$w_{\rm d}$&-1&-0.9191&-0.9461&-0.95955&-0.973&-1.044&-1.088 \\
		$\xi_2$&0&-0.1107&-0.0738&-0.05535&-0.0369&0.026095&0.05219\\
		$H_0$ (km s$^{-1}$ ${\rm Mpc}^{-1}$)&67.27&68.18&68.18&67.725&68.18&68.08&68.35\\
		\bottomrule  
	\end{tabular}
	\caption{Cosmological parameters adopted for the simulations of different models. The $\LCDM$ parameters come from the Planck 2015 results \citep{Planck2015}. The IDE1\_A and IDE2\_B ones employ the best-fitting values constrained by Planck CMB + BAO + SNIa + $H_0$ observations \citep{Costa2017}. The IDE1\_B, IDE1\_C, IDE1\_D, and IDE2\_A parameters are chosen for covering the parameter space between $\LCDM$ and the two best-fitting IDE models, see the text for a detailed description.}
	\label{tab:cosmological parameters}
\end{table*}

\begin{table*}
	\centering
	\begin{tabular}{ccccccccccccccccccc}
		\toprule
		\multicolumn{1}{c}{Model} &  \multicolumn{3}{c}{IDE1\_A} & \multicolumn{3}{c}{IDE1\_B} & \multicolumn{3}{c}{IDE1\_C} & \multicolumn{3}{c}{IDE1\_D} & \\
		\midrule
		Box size&$N_{\rm p}$&$\epsilon$&$m_{\rm p}$&$N_{\rm p}$&$\epsilon$&$m_{\rm p}$&$N_{\rm p}$&$\epsilon$&$m_{\rm p}$&$N_{\rm p}$&$\epsilon$&$m_{\rm p}$ \\
		\midrule
		100&$256^3$&4&0.045&$512^3$&6.25&0.051&$256^3$&4&0.055&$512^3$&6.25&0.057\\
		200&$512^3$&12.5&0.36&$512^3$&12.5&0.41&$512^3$&12.5&0.44&$512^3$&12.5&0.46\\
		400&$512^3$&25&2.9&$512^3$&25&3.3&$512^3$&25&3.5&$512^3$&25&3.7\\
		800&$512^3$&50&23&$512^3$&50&26&$512^3$&50&28&$512^3$&50&29\\
		\midrule
		\multicolumn{1}{c}{Model} &
		\multicolumn{3}{c}{$\LCDM$} &
		\multicolumn{3}{c}{IDE2\_A} & \multicolumn{3}{c}{IDE2\_B} \\ 
		\midrule
		Box size&$N_{\rm p}$&$\epsilon$&$m_{\rm p}$&$N_{\rm p}$&$\epsilon$&$m_{\rm p}$&$N_{\rm p}$&$\epsilon$&$m_{\rm p}$ \\
		\midrule
		100&$256^3$&4&0.065&$256^3$&4&0.067&$256^3$&4&0.070 \\
		200&$512^3$&12.5&0.52&$512^3$&12.5&0.53&$512^3$&12.5&0.56 \\
		400&$512^3$&25&4.2&$512^3$&25&4.3&$512^3$&25&4.5 \\
		800&$512^3$&50&33&$512^3$&50&34&$512^3$&50&37 \\
		\bottomrule  
	\end{tabular}
	\caption{Particle numbers $N_{\rm p}$, force softening lengths $\epsilon$ (in $h^{-1} {\rm kpc}$) and $z=0$ particle masses $m_{\rm p}$ (in $10^{10} h^{-1}\Msun$) used for the runs of different box sizes (in $h^{-1}~{\rm Mpc}$) in our $N$-body simulations.}
	\label{tab:simulation parameters}
\end{table*}

In the literature, the interaction between DE and DM is discussed by considering the energy conservation for the total dark sector. The background continuity equations of DM and DE can be written as
\begin{align}\label{eq:density}
&\dot{\rho}_{\rm c} + 3H\rho_{\rm c} = Q,\\
&\dot{\rho}_{\rm d} + 3H(1 + w_{\rm d})\rho_{\rm d} = -Q,
\end{align}
where $\rho_{\rm c}$ and $\rho_{\rm d}$ are the DM and DE energy densities respectively. $H=\dot{a}/a$ is the Hubble parameter. The dot denotes the derivative with respect to the conformal time. $w_{\rm d} = p_{\rm d}/\rho_{\rm d}$ is the equation of state for DE. $Q$ is the interaction kernel, with positive (negative) values representing energy flows from DE (DM) to DM (DE).

In phenomenological IDE models, $Q$ is usually considered as a function of DE and (or) DM densities, as well of the cosmic time, i.e. $Q=Q(\rho,H^{-1})$ (see \citealt{Wang2016} for a review). For models relating $Q$ to the DM density, \cite{Costa2017} have put tight constraints using the observations from CMB anisotropy and cosmic expansion history. However, with the same observational data sets, loose results were obtained for the models in which $Q$ is determined by the DE density only, i.e.
\begin{align}\label{eq:taylor}
Q = 3H\xi_2\rho_{\rm d},
\end{align}
where $\xi_2$ is a free parameter.
For these models, \cite{Liu2022} showed that at low redshifts when DE dominates, the non-linear structure formation in IDE models is fairly sensitive to $\xi_2$.

In this work, we focus specifically on the IDE models in the form of Equation~(\ref{eq:taylor}), and try to put tighter constraints on $\xi_2$ using the observational data from late-time structure formation.

Specifically, we will study two classes of IDE models under Equation~(\ref{eq:taylor}): IDE1 with $\xi_2<0$ and $-1<w_{\rm d}<-1/3$; and IDE2 with $\xi_2>0$ and $w_{\rm d}<-1$. The quintessence or phantom type of $w_{\rm d}$ is determined by the requirement of stable DE perturbations \citep{HE2009}. Note that in IDE1, $\xi_2<0$ so that energy transforms from DM to DE, while in IDE2 the opposite.

\subsection{Simulations}
\label{sec:2.2}
As the interaction between DE and DM introduces fundamental differences compared to a standard $\LCDM$ cosmology, careful modifications are required for simulating the evolution of an IDE universe. To generate the simulation initial conditions, we first apply the capacity constrained Voronoi tessellation (CCVT) method \citep{Liao2018,Zhang2021} to produce a uniform and isotropic particle distribution. Compared to the gravitational equilibrium glass distribution \citep{White1996}, the CCVT distribution is a geometrical equilibrium state and is a more natural choice for models considering not just pure gravitational interaction.
We then modify the \textsc{camb} code consistently to calculate the initial matter power spectra \citep{Lewis2002}, and use our modified \textsc{2lptic} code to generate the perturbations on particle positions and velocities \citep{Crocce2006}.
To perform IDE simulations, we employ the fully self-consistent code \textsc{me-gadget} \citep{Zhang2018}, which is modified from \textsc{gadget-2} \citep{Springel2005} by including the consistent calculations of expansion history $H(a)$, of particle mass evolution, of Poisson equation and Euler equation in IDE models. See \citet{Zhang2018} for detailed descriptions of the modifications above.

We aim at exploring the parameter space allowed by the constraints of \cite{Costa2017} using Planck CMB + BOSS BAO + SNIa + $H_0$ observations. The three most important parameters related are the DE-DM interaction parameter $\xi_2$, the matter density $\Omega_{\rm m}$ and the equation of state of DE $w_{\rm d}$, which are highly degenerated with each other for the IDE1 and IDE2 models with the data considered by \cite{Costa2017} (see their Figure~4). Accordingly, we performed 7 sets of simulations, by varying these parameters linearly,  to cover the parameter space, named IDE1\_A, IDE1\_B, IDE1\_C, IDE1\_D, $\LCDM$, IDE2\_A, and IDE2\_B, in the sequence of the change of parameter values. In detail, the IDE1\_A and IDE2\_B correspond to the best-fitting parameters of IDE1 and IDE2 models in \cite{Costa2017}, respectively. The IDE1\_C employs the mid-point values between $\LCDM$ and IDE1\_A, while the IDE2\_A one takes the mid-point values between $\LCDM$ and IDE1\_B. Lastly, the IDE1\_B and IDE1\_D ones are located at the two tertile positions between $\LCDM$ and IDE1\_A. The parameters of the above models are listed in Table~\ref{tab:cosmological parameters}, all the chosen 
parameters are well within the allowed range of the \cite{Costa2017} constraints.

For each model, we performed 4 simulation runs with box sizes of $\{$100, 200, 400, 800$\}$ $h^{-1}$Mpc, in order to investigate different mass scales. Each run evolves $256^3$ particles, except the 100 $h^{-1}$Mpc runs for IDE1\_A, IDE1\_C, $\LCDM$, IDE2\_A and IDE2\_B with $512^3$ particles. The force softening length is set as ${\sim} 1/50$ of the mean separation of simulated particles. We summarise the simulation information in Table~\ref{tab:simulation parameters}. For the runs with the same box size, we use the same random seed to produce initial conditions so as to avoid cosmic variance. To match the redshift distribution of our observation data (see details in Section~\ref{sec:3.1}), we output the simulation snapshots at redshift $z = \left\{\text{0.0, 0.1, 0.2, 0.3, 0.4, 0.5, 0.6, 0.7, 0.8, 0.9, 1.0}\right\}$ for each run.

The DM haloes are identified by the Amiga Halo Finder \citep[\textsc{ahf},][]{Knollmann2009}, which has been consistently modified for taking into account the particular evolution of $\Omega_{\rm m}$ due to DE-DM interactions. We employ the centre of mass of the particle group as the halo centre, and define the virial radius, $r_{200}$, as the radius within which the mean density is 200 times the cosmic critical density $\rho_{\rm crit}(z)$. In order to study the halo internal structure accurately, we only select the haloes with $\geq$ 1000 member particles to ensure that there are enough particles when measuring the halo density profile (see details in Section~\ref{sec:3.1}). Combined the multiple box sizes for each model, our halo catalogs contain well-resolved DM haloes in a widely covered mass range of $5\times10^{11} h^{-1} \Msun <M_{200}<10^{15} h^{-1} \Msun$, and keep a rich number of samples at the same time.

\section{Halo concentration--mass relations from simulations}
\label{sec:3}
In the bottom-up hierarchical clustering scenario of cold dark matter, haloes formed from the peaks of initial cosmic density perturbations and grew on mass by accreting environmental matter and merging smaller objects. In this picture, the radial density distribution of DM haloes was shaped by their cosmic formation histories which depend essentially on the underlying cosmological models. Cosmological $N$-body simulations have proved that haloes exhibit a scale-free self-similarity, i.e. the radial density profile of haloes of all mass ranges can be well described by the Navarro--Frenk--White \citep[NFW,][]{Navarro1996} functional profile
\begin{align}\label{eq:NFW}
\rho(r) = \frac{\rho_{\rm s}}{(r/r_{\rm s}) (1 + r/r_{\rm s})^2},
\end{align}
where $\rho_{\rm s}$ is the characteristic density, and $r_{\rm s}$ is the scale radius where the logarithmic slope $\rm{d}$$\ln \rho(r)/\rm{d}$$\ln r = -2$. The ratio between the virial radius $r_{200}$ and this scale radius $r_{\rm s}$ characterizes the halo concentration
\begin{align}\label{eq:c200}
c_{200} = \frac{r_{200}}{r_{\rm s}}.
\end{align}
Equivalently, one can rewrite the NFW profile and replace the two free parameters $\rho_{\rm s}$ and $r_{\rm s}$ with the halo concentration $c_{200}$ and halo virial mass $M_{200}$. Given the different mass assembly histories of the haloes with different present masses, the relation between $c_{200}$ and $M_{200}$ ($c$--$M$ relation in short) reveals profound cosmological information, and this relation has been 
studied extensively in alternative models
\citep[e.g.][]{Peter2010,Barreira2014,Ludlow2016}.

In order to put constraints on IDE models with the observational $c$--$M$ relations at different redshifts, we first compute and fit the $c$--$M$ relations based on our IDE simulations, then analyse the redshift evolution, and study the dependence on the interaction parameter $\xi_2$.

\begin{figure*}
	\includegraphics[width=\textwidth]{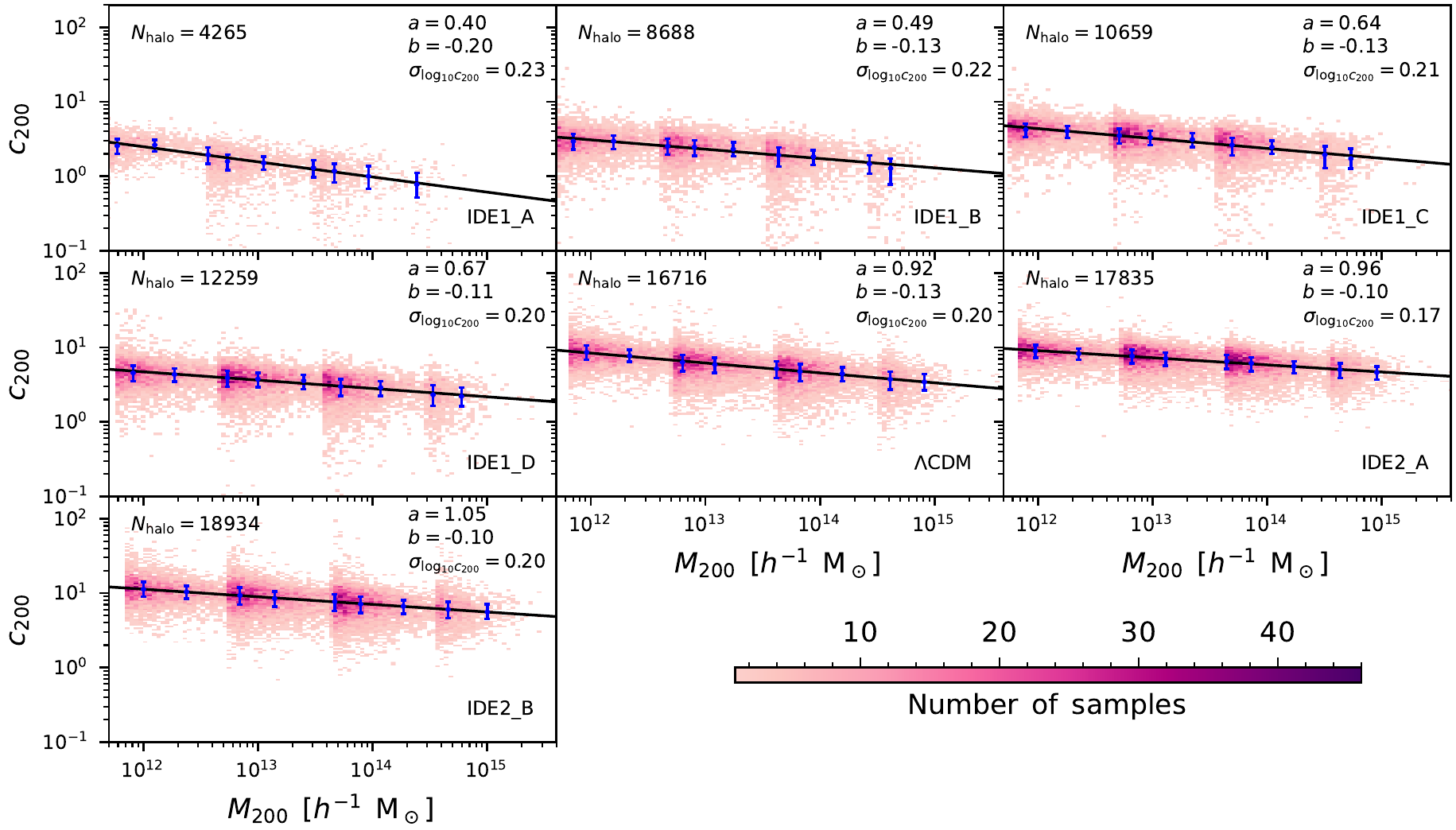}
 \caption{Halo concentration--mass relation in our simulations at $z=0$. The coloured pixels present the distribution of the haloes and the blue bars show the median concentration and $1\sigma$ scatter in each mass bin. Note that for each model, we have considered the haloes from all four simulations with different box sizes. The solid lines show the best-fitting $c$--$M$ relation. The best-fitting values of parameters $a$ and $b$, and the average scatter of halo concentrations are given in the top-right corner of each panel. The total halo numbers in different models are given in the top-left corner of each panel.}
	\label{fig:cmrelation}
\end{figure*}

\subsection{Computing and fitting the $c$--$M$ relations}
\label{sec:3.1}
For fitting the simulated haloes with the NFW profile, we calculate the spherically averaged densities of each halo in 15 radial shells which are binned logarithmically from three times the softening length to $r_{200}$. Then, we search for the best-fitting profile by minimizing the figure-of-merit function
\citep{Navarro2010}
\begin{align}\label{eq:Q2}
\chi^2 = \sum\limits_{i=1}\limits^{N_{\rm bins}}(\ln{\rho_i}-\ln{\rho_i^{\rm NFW}})^2,
\end{align}
where $\rho_i$ is the spherically averaged density in the \textit{i}th spherical shell, and $\rho_i^{\rm NFW}$ is the corresponding density calculated by Equation~(\ref{eq:NFW}) by taking $\rho_{\rm s}$ and $r_{\rm s}$ as free parameters.

In Fig.~\ref{fig:cmrelation}, we plot all $z=0$ haloes on the $c$--$M$ plane for different models. 
Note that at $z=0$, we have $> 4000$ haloes in the IDE1\_A model which has the minimum halo number, and $> 16000$ haloes in the $\LCDM$ and IDE2 models (see the top-left corner of each panel for the detailed halo numbers in different models). Such large halo samples allow us to have a robust measurement and error estimation of the $c$--$M$ relation. To measure the $c$--$M$ relation, we divide haloes into different mass bins, and compute the median value and the standard deviation in each bin (the blue data points in Fig.~\ref{fig:cmrelation}). 
Generally, the halo concentration from all models considered here has a negative correlation with the halo mass, which reflects a common bottom-up structure formation paradigm that haloes with lower present-day masses formed at an earlier epoch when the cosmic matter density was higher. For all models in Fig.~\ref{fig:cmrelation}, we find that their $c$--$M$ relations can be well fitted with a power-law form
\begin{align}\label{eq:cM}
\log_{10} c_{200} = a + b \log_{10} (M_{200}/[10^{12} h^{-1} \Msun]),
\end{align}
where $b$ is the logarithmic slope, and $a$ is the intercept of $\log_{10} c_{200}$ at $M_{200}=10^{12} h^{-1} \Msun$. 
The solid lines in Fig.~\ref{fig:cmrelation} show the best-fitting $c$--$M$ relations, and the corresponding fitted parameters $a$ and $b$ are summarised in each panel. Note that when performing the fitting, we use the median values in different mass bins and adopt their $1\sigma$ errors as weights. The intercept $a$ shows a clear dependence on the models. At $z=0$, the $a$ in IDE1 models are all smaller than that of the $\LCDM$ model, while the two IDE2 models have the largest $a$. On the other hand, the $c$--$M$ relations from all models are nearly parallel, indicating a similar value of slope $b$. These results are consistent with the conclusions in \cite{Liu2022}. That is, in IDE1 in which DM decays into DE, haloes have both smaller masses and lower concentrations than their $\LCDM$ counterparts, which shifts the $c$--$M$ relation towards the left and bottom sides in Fig.~\ref{fig:cmrelation}. Both effects lead to a lower value of $a$ in IDE1 models. Meanwhile, the opposite cases occur in IDE2 models with DE transforming into DM.

\subsection{Redshift evolution of the $c$--$M$ relations}\label{subsec:c_M_relation_redshift}
\label{sec:3.2}

\begin{figure}
	\includegraphics[width=\columnwidth]{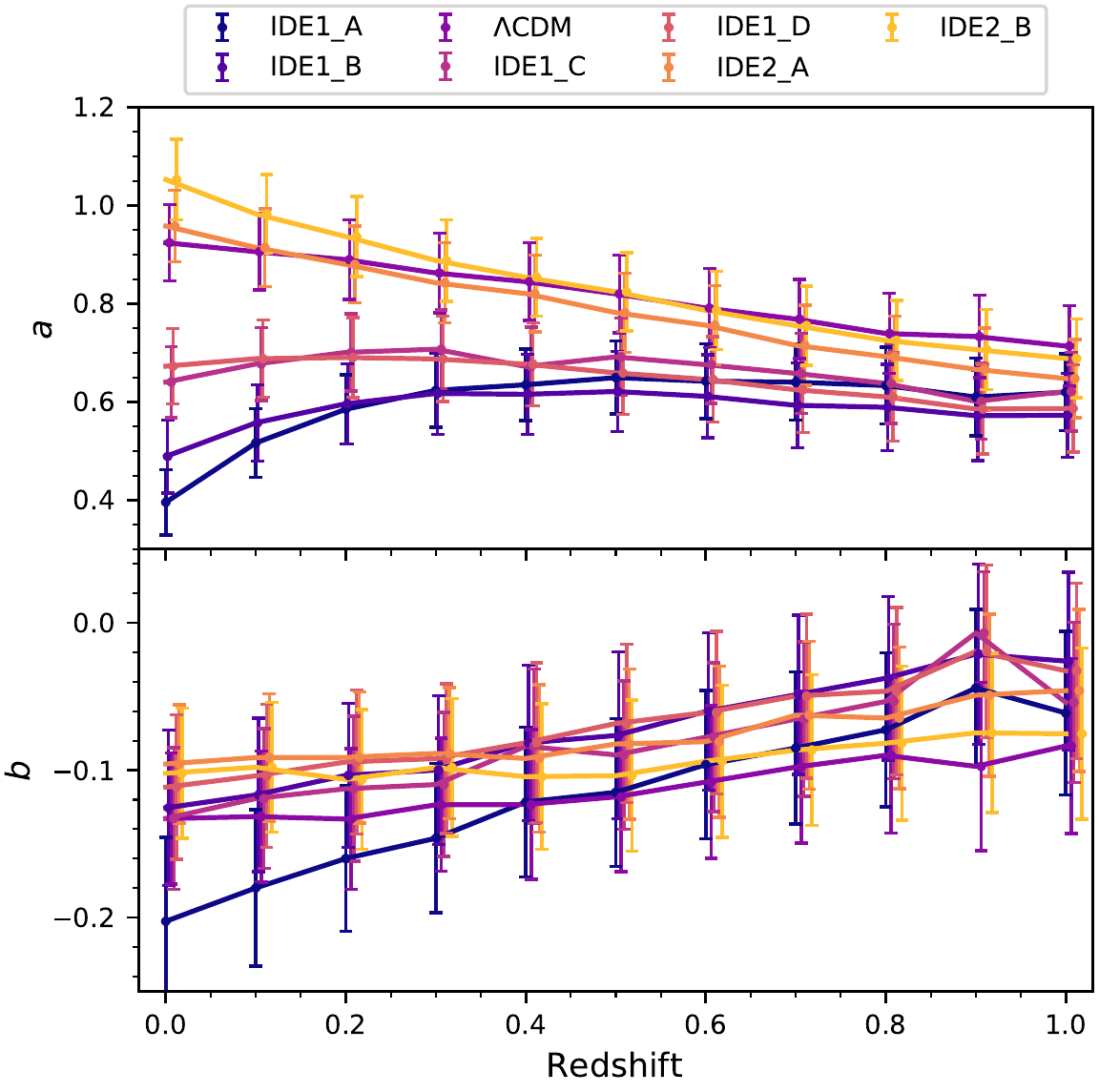}
	\caption{The redshift evolution of $c$--$M$ relation parameters in different models. The top panel shows the results of intercept $a$, while the bottom one for the logarithmic slope $b$. The solid lines connect the best-fitting values, and the error bars show 1$\sigma$ uncertainties. Note that for each redshift, the data points from different models have been slightly shifted for a clearer illustration.}
	\label{fig:abfit}
\end{figure}

Similar to the $\Lambda$CDM model \citep[e.g.][]{Dutton2014}, at higher redshifts, the $c$--$M$ relations of IDE models can also be well fitted by Equation~(\ref{eq:cM}). We first discuss the slope parameter $b$. Essentially, the absolute value of $b$ reflects the magnitude of differences between low-mass and high-mass haloes on their formation histories. The evolution of $b$ from different models is shown in the bottom panel of Fig.~\ref{fig:abfit}. Generally, $b$ is negative at $z<1$ and keeps decreasing at lower redshifts, showing the accumulative discrepancies of halo formation across mass scales. For IDE models, interestingly, it seems that $b$ decreases faster in the models with smaller $\xi_2$ (i.e. IDE1), but this trend is very weak. At each redshift, within the $1\sigma$ uncertainty, there is no evident difference in $b$ among different models, reflecting that the $c$--$M$ relations from different models are almost parallel as shown in Fig.~\ref{fig:cmrelation} for the $z =0 $ case.

Given the similar slope $b$ at every particular redshift, the offset of intercept parameter $a$ reveals the systematic contrasts between models that contributed from both halo concentrations and halo masses. In the top panel of Fig.~\ref{fig:abfit} for the evolution of $a$, we can see prominent differences among models, especially at low redshifts. It is obvious that the values of $a$ evolve in two directions: 
in $\LCDM$ and IDE2 models, $a$ keeps growing as redshift decreases from $z=1$ to $0$, whereas in IDE1 models, $a$ first grows slightly and turns around at $z \sim 0.4$ and keeps decreasing. This demonstrates that the DE-DM interaction gives rise to continuous changes in halo structures and, significantly, this effect is expected to be detected more signally at the late epoch. Reasonably, the differences of $a$ among models are consistent with their interaction parameter $\xi_2$. We analyse this correlation quantitatively in the following subsection.

\subsection{Dependence of the $c$--$M$ relations on $\xi_2$}
\label{sec:3.3}

\begin{figure}
    \centering
	\includegraphics[width=\columnwidth]{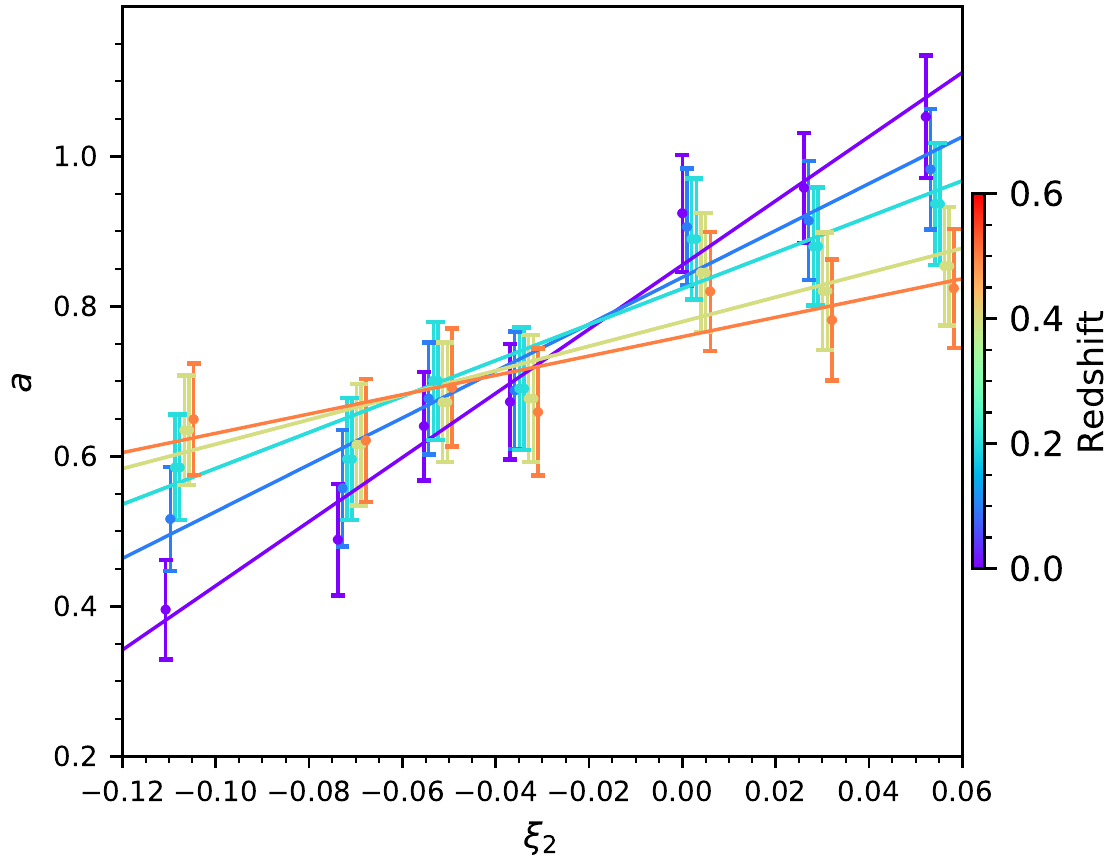}
	\caption{The relation between intercept parameter $a$ and DE-DM interaction parameter $\xi_2$ at different redshifts. The data points with error bars show the best-fitting values of $a$ and the 1$\sigma$ uncertainties. The solid lines present the best-fitting results of this relation with $a=m_a\xi_2+n_a$. For a particular model (i.e. a particular $\xi_2$), the data points from different redshifts have been slightly shifted for a better illustration.}
	\label{fig:axi2}
\end{figure}

\begin{table}
	\centering
	\begin{tabular}{cccc}
		\toprule  
		Redshift $z$&$m_a(z)$&$n_a(z)$&$n_b(z)$\\
		\midrule  
		0&4.2776&0.8549&-0.1237\\
        0.1&3.1235&0.8385&-0.1174\\
        0.2&2.3969&0.8234&-0.1137\\
        0.3&2.3969&0.8234&-0.1137\\
        0.4&1.6332&0.7794&-0.0989\\
        0.5&1.6332&0.7794&-0.0989\\
        0.6&1.2882&0.7594&-0.0943\\
		\bottomrule  
    \end{tabular}
    \caption{The best-fitting results of Equation~(\ref{eq:a-xi2}) and (\ref{eq:b-xi2}) which model the relation between $c$--$M$ relation parameters and DE-DM interaction parameter $\xi_2$.}
    \label{tab:mn}
\end{table}

In \cite{Liu2022}, they found that the ratio of $c_{200}$ between the halo counterparts in IDE and $\LCDM$ models is proportional to the interaction parameter $\xi_2$, and is independent of halo mass. Following the clues revealed by \cite{Liu2022}, here we quantify the correlation between the $c$--$M$ relation fitted parameters (i.e. $a$ and $b$) and $\xi_2$ at different redshifts.

Fig.~\ref{fig:axi2} shows the relation between parameter $a$ and $\xi_2$, as well as its dependence on redshift. The data points with error bars are the best-fitting values and $\pm 1 \sigma$ errors of $a$ obtained from different models and different redshifts. Given a redshift, the correlation between $a$ and $\xi_2$ can be well fitted by a simple linear model
\begin{align}\label{eq:a-xi2}
a(\xi_2, z) = m_a(z)\xi_2+n_a(z),
\end{align}
as plotted by the solid lines. It is worth noting that the slope, $m_a$, of this linear model increases with decreasing redshift, implying that the effects from DE-DM interaction become more prominent at the later epoch of the Universe.

As discussed in Section~\ref{subsec:c_M_relation_redshift}, at each redshift, there is no clear difference of $b(z)$ among different models within the $1\sigma$ uncertainty, therefore, we choose a $\xi_2$-independent model for the logarithmic slope
\begin{align}\label{eq:b-xi2}
b(z) = n_b(z),
\end{align}
where $n_b(z)$ is the mean of $b$ from different models at redshift $z$.

The values of $m_a(z)$, $n_a(z)$ and $n_b(z)$ calculated from simulation snapshots at different reshifts are summarised in Table~\ref{tab:mn}.

\section{Observational Constraints}
\label{sec:4}
\subsection{Observational Data}
\label{sec:4.1}

In the previous section, we have calculated the $c$--$M$ relation from $N$-body simulations and provided the parametrized form which depends on the redshift and interaction parameter $\xi_2$. Therefore, we may use observational data to put constraints on the IDE models. 

There are several ways of measuring the $c$--$M$ relation from observations, such as using the X-ray data \citep{Amodeo2016, Ettori2010, Comerford2007, Groener2016}, the gravitational lensing \citep{Merten2015, Meneghetti2014, Sereno2010, Mandelbaum2006, Comerford2007, Groener2016}, and the galaxy rotation curve \citep{Martinsson2013}. We select the results with which the concentration and mass definition are consistent with our simulations.

\begin{table*}
    \centering
    \begin{tabular}{|c|c|c|c|c|}
    \hline
        Redshift range & Mass range $(h^{-1}\Msun)$ & Number of samples & Method & Reference \\ \hline
        $[0.405, 0.64]$ & [3.14e+14, 5.73e+15] & 27 & X-ray &\cite{Amodeo2016} \\
        $[0.009, 0.081]$ & [1.93e+13, 2.20e+14] & 15 & X-ray &\cite{Gastaldello2007} \\
        $[0.092, 0.307]$ & [1.34e+14, 3.53e+15] & 44 & X-ray &\cite{Ettori2010} \\
        $[0.003, 0.64]$ & [1.93e+13, 1.86e+16] & 649 & Gravitational lensing, X-ray &\cite{Groener2016} \\
        $[0.003, 0.62]$ & [5.57e+13, 4.57e+15] & 60 & Gravitational lensing &\cite{Comerford2007} \\
        $[0.188, 0.545]$ & [4.60e+14, 1.42e+15] & 17 & Gravitational lensing &\cite{Merten2015} \\
        $[0.188, 0.545]$ & [5.71e+14, 1.15e+15] & 18 & Gravitational lensing &\cite{Meneghetti2014} \\
        $[0.28, 0.637]$ & [2.86e+14, 1.71e+15] & 4 & Gravitational lensing &\cite{Sereno2010} \\
        0.36 & [4.14e+13, 9.57e+13] & 2 & Gravitational lensing &\cite{Mandelbaum2006} \\
        $\approx 0$ & [1.99e+11, 4.86e+12] & 30 & Fitting rotation curves & \cite{Martinsson2013} \\ \hline
    \end{tabular}
    \caption{Observational $c$--$M$ relation data used in this study. From left to right, we show the redshift range, mass range, number of samples, observational method, and the corresponding reference for each data set.}
	\label{tab:obser_data}
\end{table*}

\begin{figure*}
	\includegraphics[width=\textwidth]{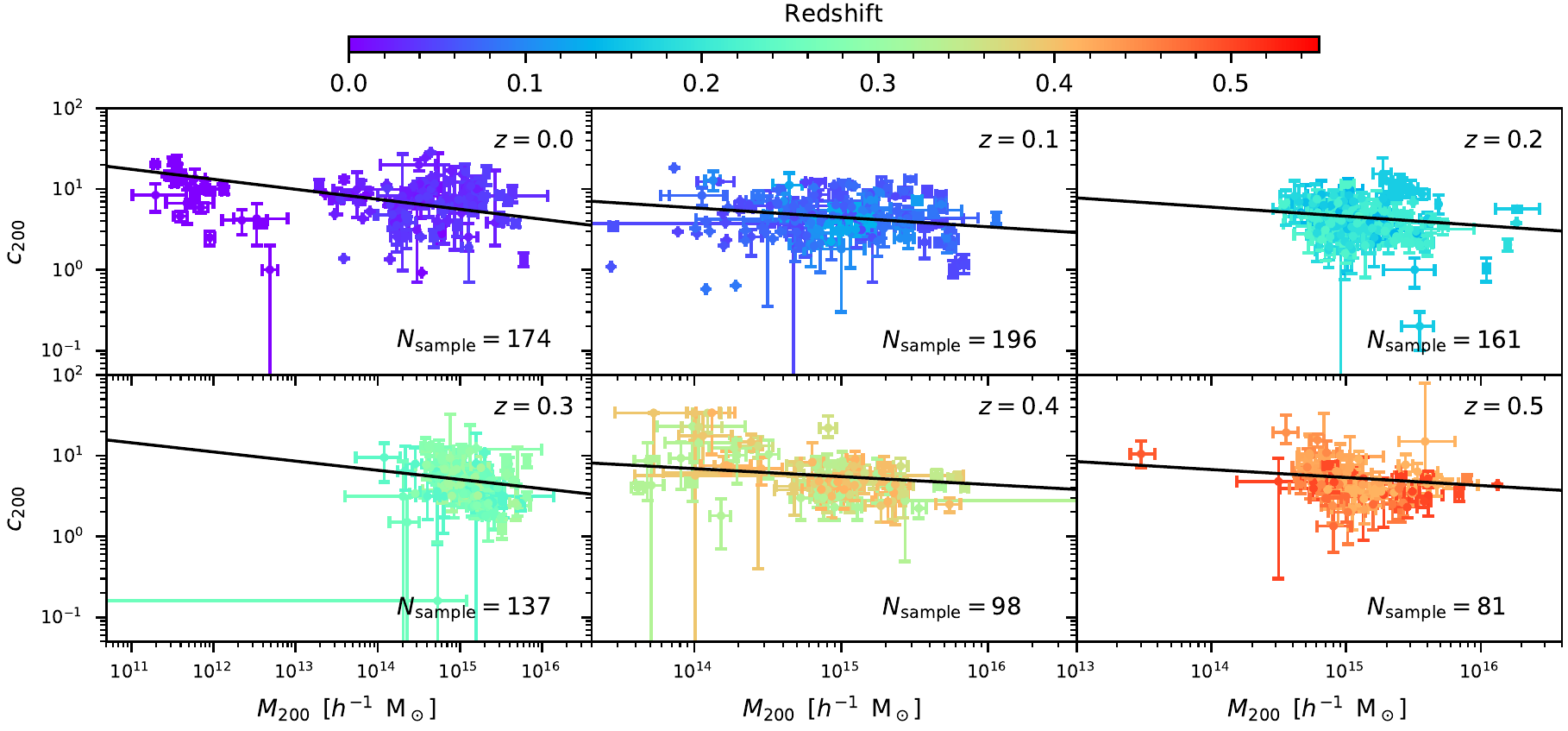}
	\caption{Observational $c$--$M$ relation data binned by redshift. The data points with error bars show the observational samples described in Section~\ref{sec:3.1} and Table~\ref{tab:obser_data}, with different colours indicating the redshift. The black lines show the best-fitting results with IDE models and the corresponding values of $\xi_2$ are listed in Table~\ref{tab:likelihood}.}
	\label{fig:obser}
\end{figure*}

For the X-ray data, the image is first obtained from X-ray observatories. Then the temperature profile can be estimated based on the images of different energy. Using the hydrostatic equilibrium assumption, the density profile can also be calculated. The NFW density profile fitting can be applied to the estimated density profile to get the mass and concentration of each galaxy group. Alternatively, after forward modelling the temperature profile from the NFW density profile, by fitting it to the observed temperature profile, the mass and concentration of each galaxy group can be obtained. The errors come from both observational data and models. 

The gravitational lensing signal comes from the distortion of the light from background galaxies due to the foreground lens galaxies. Such distortion is caused by the gravity of the lens galaxies, therefore, gravitational lensing can directly indicate the mass distribution. The NFW profile parameterization of the lens galaxies is constrained by the lensing signal, the mass and concentration can then be measured.

The galaxy rotation curve measurement can only be applied to low-redshift galaxies since we need to have a high-resolution spectral observation to measure the rotation of galaxies accurately. Assuming the NFW profile for the galaxies, the rotation curve can be calculated. So by fitting to the observed rotation curve, the mass and concentration can be obtained.

The overall description of the observational data is provided in Table~\ref{tab:obser_data}. In the table, we have collected the $c$--$M$ relation measurement from literature and included the information of the redshift range, mass range, number of samples, and observational methods. The data samples include the concentration and mass of each halo measured from observations and the error bar of each sample. We have followed \citet{Audi2017} to estimate the equivalent symmetric errors from the asymmetric errors from the data samples, then we have calculated the error in the logarithmic scale for our fitting purpose.

We further rearranged the data samples into redshift range $z=0.0$--$0.05$, $0.05$--$0.15$, $0.15$--$0.25$, $0.25$--$0.35$, $0.35$--$0.45$, $0.45$--$0.55$, $0.55$--$0.65$, which are labeled as $z=0, 0.1, 0.2, 0.3, 0.4, 0.5, 0.6$ separately. The data we have used for constraints and the fitted curves are shown in Fig. \ref{fig:obser}. The number of samples in each redshift bin is shown in the figures as well.

\subsection{Likelihood}
\label{sec:4.2}

\begin{figure}
	\includegraphics[width=\columnwidth]{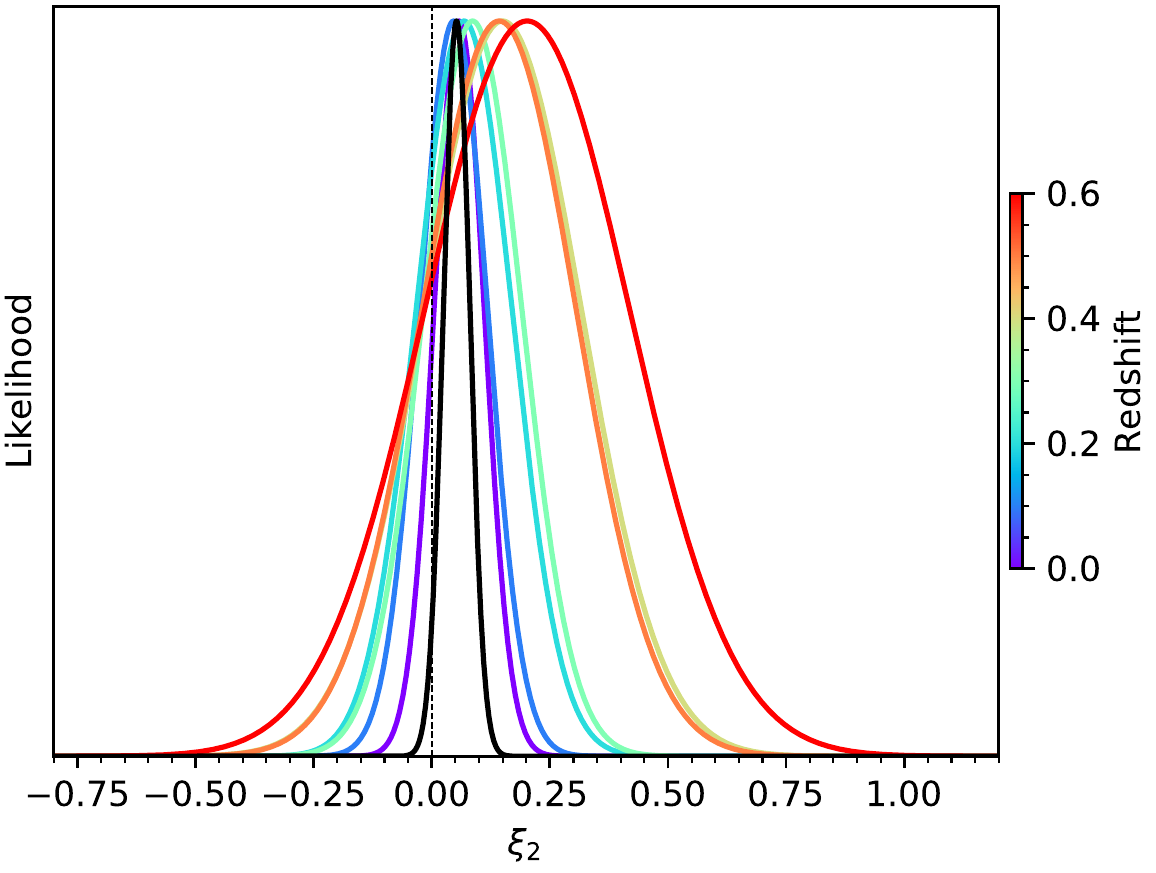}
	\caption{The normalized likelihood distribution for DE-DM interaction parameter $\xi_2$. The coloured lines show the results from different redshift bins, while the black line represents the combined results. The best-fitting values and $1\sigma$ uncertainties of $\xi_2$ are listed in Table~\ref{tab:likelihood}.}
	\label{fig:likelihood}
\end{figure}

In order to find the best-fitting value of the interaction parameter $\xi_2$ and the corresponding error, we have used the maximum likelihood method. Our likelihood function at each different redshift is constructed as
\begin{align}\label{eq:likelihood}
L = \exp{\left\{-\frac{1}{2}\sum\limits_{i}\frac{(\log_{10}c_{200}^{\rm obs.}-\log_{10}c_{200}^{\rm mod.})^2}{\sigma_{\log_{10}c_{200}^{\rm obs.}}^2+\sigma_{\log_{10}M_{200}^{\rm obs.}/[10^{12} h^{-1} \Msun]}^2+\sigma_{\log_{10}c_{200}^{\rm sim.}}^2}\right\}},
\end{align}
where the concentration parameter $c_{200}^{\rm mod.}$ is related to the halo mass $M_{200}$ using our simulations, and the formula is  
\begin{align}\label{eq:c200modelb}
\log_{10}c_{200}^{\rm mod.} = m_a(z)\xi_2 + n_a(z) + n_b(z) \log_{10} (M_{200}/[10^{12} h^{-1} \Msun]),
\end{align}
where the parameter $m_a$, $n_a$, and $n_b$ are fixed by our simulations, as a function of redshift. 

It can be seen that the key of our likelihood is the estimation of errors. The errors can be separated into three parts, the error of concentration from observational data, the error of mass from observations, and the error of concentration representing the intrinsic scatter of the $c$--$M$ relation. The observational error is provided by the references. For the intrinsic scatter, we have estimated the level by calculating the standard deviation of the $c$--$M$ relation using our simulations, which is shown in Fig.~\ref{fig:cmrelation}. We have found an average level of $\sigma_{\log_{10}c_{200}^{\rm sim.}}=0.2\,{\rm dex}$. By taking the degrees of freedom and normalization of likelihood into account, we illustrate the likelihood function in Fig.~\ref{fig:likelihood}. 

\subsection{Constraints}
\label{sec:4.3}

\begin{table}
	\centering
	\begin{tabular}{cc}
		\toprule  
		Redshift $z$& Best-fitting $\xi_2$ \\ 
		\midrule  
		0&$0.058\pm0.053$\\
		0.1&$0.046\pm0.071$\\
		0.2&$0.068\pm0.101$\\
		0.3&$0.086\pm0.104$\\
		0.4&$0.150\pm0.166$\\
		0.5&$0.143\pm0.162$\\
		0.6&$0.202\pm0.216$\\
		Combined redshifts&$0.071\pm0.034$\\
		\bottomrule  
	\end{tabular}
	\caption{The best-fitting values and $1\sigma$ uncertainties for the DM-DE interaction parameter $\xi_2$ constrained from different redshifts. The bottom row shows the result obtained by combining all redshift bins.}
	\label{tab:likelihood}
\end{table}

Because we have fixed the parametrized form of the $c$--$M$ relation using simulations, the likelihood is actually only a function of the DM-DE interaction parameter $\xi_2$. By maximizing the likelihood function we have found the best-fitting $\xi_2$ and its error in each different redshift bin. After being weighted by the degrees of freedom and the number of data points, the likelihood of seven redshift bins can be multiplied together. Combining the likelihood of seven redshift bins, we have obtained the likelihood function in black in Fig.~\ref{fig:likelihood}, the best-fitting value and the error range are shown in Table~\ref{tab:likelihood}. The best-fitting $c$--$M$ relation is plotted using the black curves at different redshifts in Fig.~\ref{fig:obser}.

We can see that the higher the redshift, the wider the likelihood function, because we have less observational data points at higher redshifts, and also because the errors of the observational data at higher redshift are larger. There seems to be a slight evolution of the best-fitting $\xi_2$ value with redshift, but the significance is low and we cannot make a conclusion about it. This might be caused by the selection effects of the observational $c$--$M$ relations and it will be interesting to investigate once we have enough data.   

In conclusion, we have found that by combining seven redshift bins, the best-fitting results are $\xi_2=0.071\pm0.034$. This indicates that we have found a preference for DE transferring into DM at late time universe, at the significance level of $2.1\sigma$. However, without further investigation into baryonic effects \citep{Gnedin2004,Schaller2015,Desmond2017,Anbajagane2022} on the $c$--$M$ relation, it is too early to conclude with any strong evidence. The $\LCDM$ model is still within the $2.1\sigma$ range. Our conclusion is consistent with \citet{Zhang2019}, which used galaxy-galaxy lensing for the constraints. 

\section{Conclusion and Discussion}
\label{sec:5}

The $c$--$M$ relation is one of the key relations about dark matter haloes and it encodes the halo assembly history. Comparing the $c$--$M$ relation from theoretical predictions to the observations can help us extract useful cosmological information and put constraints on the IDE models. This idea has been studied preliminarily in \citet{Liu2022}, which shows that the $c$--$M$ relation is potentially a simple and powerful probe for the DM-DE interaction parameter. 

In this study, we have employed a large set of cosmological $N$-body simulations using the \textsc{me-gadget} code \citep{Zhang2018} to further explore the $c$--$M$ relations in the IDE models. Following \citet{Zhang2019}, we have performed simulations which cover wider IDE parameter space and have a wider range of resolution. We have measured the IDE $c$--$M$ relations from the simulations and found that they can be well fitted by a simple power law. We also provided simple fitting formulae for the parameters as functions of redshift $z$ and the IDE parameter $\xi_2$. In our simulations, the intrinsic scatter of the modelled $c$--$M$ relations is found to be about 0.2 dex for the concentration.

In our fitting formulae, the interaction parameter $\xi_2$ and redshift $z$ are the only two parameters that affect our $c$--$M$ relations. Then we have used the observed $c$--$M$ relation from the literature to put constraints on the interaction parameter $\xi_2$. We have constructed the likelihood function and used the maximum likelihood method to find the best-fitting $\xi_2$ at different redshift bins. By combining seven redshift bins, we have obtained a tight constraint on the DM-DE interaction parameter $\xi_2=0.071\pm0.034$.

We summarise our results in the following points:
\begin{itemize}
    \item We have used $N$-body simulations with different resolutions to illustrate that the $c$--$M$ relations can be well fitted by power laws in the IDE model. We have provided the fitting formulae in Equation~(\ref{eq:cM}) and listed the best-fitting parameters in Fig.~\ref{fig:cmrelation}.
    \item Using the simulations representing different IDE models, we have found the $c$--$M$ relation fitting parameters as a function of the interaction parameter $\xi_2$, following Equation~(\ref{eq:c200modelb}), and the fitting results are shown in Table~\ref{tab:mn}.
    \item We have used the observed $c$--$M$ relation to constrain the interaction parameter $\xi_2$, as shown in Fig.~\ref{fig:obser}.
    \item Using the maximum likelihood method, comparing to the observation data, we have found that $\xi_2=0.071\pm0.034$. This means our results favor dark energy transferring into the dark matter in the late universe at a $2.1\sigma$ level.  
\end{itemize}

As a first attempt, we have found an interesting positive indication that dark energy transfers into dark matter in the late universe from the observed halo $c$--$M$ relation. In our future studies, in order to build up a more accurate model, we will construct large sets of hydrodynamical simulations with baryonic effects \citep{Gnedin2004,Schaller2015,Desmond2017,Anbajagane2022} properly included, explore larger cosmological parameter space with more simulations, and carry out more detailed studies about different effects, such as selection bias in the $c$-$M$ relation from different measurement methods and observational data sets, and projection effects, which could introduce a ${\sim} 10$ per cent bias as shown in \citet{Du2015}. This newly build-up model will thus be further applied to future larger observational data samples to better constrain the interaction strength between dark matter and dark energy.

\section*{Acknowledgements}

YL acknowledges supports from the National Key Research and Development of China (No.2018YFA0404503), the NSFC grants (No.12033008, No.11988101), the K.C.Wong Education Foundation, and the science research grants of the China Manned Space Project (No.CMS-CSST-2021-A03 and No.CMS-CSST-2021-A07). SL acknowledges the support by the European Research Council via ERC Consolidator Grant KETJU (no. 818930). JZ acknowledges the support from the China Manned Space Project with no. CMS-CSST-2021-A03. XKL acknowledges support from the NSFC of China under grant No. 11803028 and No. 12173033, YNU grant No. C176220100008, and the research grants from the China Manned Space Project No. CMS-CSST-2021-B01. WD acknowledges the support from NSFC grants 11803043, 11890691 and 11720101004. The calculations of this study were partly done on the Yunnan University Astronomy Supercomputer.

We are thankful to the community developing and maintaining software packages extensively used in our work, including \textsc{Matplotlib}~\citep{Hunter2007}, \textsc{NumPy}~\citep{Harris2020}, and \textsc{SciPy}~\citep{2020SciPy}.

\section*{Data Availability}

The simulation data underlying this article will be shared on a reasonable request to the authors.
 



\bibliographystyle{mnras}
\bibliography{biblio}








\bsp	
\label{lastpage}
\end{document}